\documentclass[prd,aps,showpacs,,nofootinbib,superscriptaddress,preprintnumbers,epsf,psf]{revtex4}

\usepackage[english]{babel}
\usepackage{pstricks}
\usepackage[dvips]{graphicx}
\usepackage{epsf}
\usepackage{amsmath}
\usepackage{amsfonts}
\usepackage{amssymb}

\begin{document}

\title{Asymmetries associated with higher twists:
gauge invariance, gluonic poles and twist three}

\author{\underline{I.~V.~Anikin}}
\email{anikin@theor.jinr.ru}
\affiliation{Bogoliubov Laboratory of Theoretical Physics, JINR,
             141980 Dubna, Russia}
\affiliation{Institute for Theoretical Physics, University of Regensburg,
             D-93040 Regensburg, Germany}
\author{O.~V.~Teryaev}
\email{teryaev@theor.jinr.ru}
\affiliation{Bogoliubov Laboratory of Theoretical Physics, JINR,
            141980 Dubna, Russia}

\begin{abstract}
We explore the electromagnetic gauge invariance of the hadron tensor of
the Drell-Yan process with one transversely polarized hadron.
Due to the special role of the contour gauge for gluon fields,
the prescription for the gluonic pole in the twist $3$ correlator can be
related to the causality prescriptions for exclusive hard processes.
Because of this, we find the extra contributions, which naively
do not have an imaginary phase. The single spin asymmetry for
the Drell-Yan process is accordingly enhanced by the factor of two.
\end{abstract}
\date{\today}
\maketitle

\textbf{Introduction}. The problem of the electromagnetic gauge invariance in the
deeply virtual Compton scattering (DVCS) and similar exclusive processes
has intensively been discussed during last few years, see for example
\cite{Gui98, Pire, APT-GI, Bel-Mul, PPSS, MB}.
This development explored the similarity with the earlier studied inclusive spin-dependent
processes \cite{Efr}, and the transverse component of momentum transfer in DVCS
corresponds to the transverse spin in DIS.
Here we combine the different approaches to apply them in the
relevant case of the Drell-Yan (DY) process where one of hadrons is the transversally
polarized nucleon.
The source of the imaginary part, when one
calculates the single spin asymmetry associated with
the DY process, is the quark propagator in
the diagrams with quark-gluon (twist three) correlators. This leads \cite{Teryaev,Boer} to
the gluonic pole contribution to SSA.
The reason is that these boundary conditions provide the purely real
quark-gluon function $B^V(x_1,x_2)$ which parameterizes
$\langle\bar\psi\gamma^+A_\alpha^T\psi\rangle$ matrix element.
By this fact the diagrams with two-particle correlators do not contribute to the
imaginary part of the hadron tensor related to the SSA \cite{Ter00}.
In our paper, we perform a thorough analysis of the transverse polarized DY hadron tensor
in the light of the QED gauge invariance, the causality and gluonic pole contributions.
We show that,
in contrast to the naive assumption,
our new-found additional contribution is directly related to
the certain complex prescription in the gluonic pole $1/(x_1-x_2)$ of
the quark-gluon function $B^V(x_1,x_2)$ (cf. \cite{BQ} and
see e.g.\cite{c} and Refs. therein).
Finally,
the account for this extra contributions corrects  the SSA formula for
the transverse polarized Drell-Yan process by the factor of $2$.
Note that
our analysis is also important in view of the recent investigation
of DY process within both the
collinear and the transverse-momentum factorization schemes  with
hadrons replaced by on-shell parton states  \cite{Cao:2009rq}.

\begin{figure}[t]
\centerline{\includegraphics[width=0.3\textwidth]{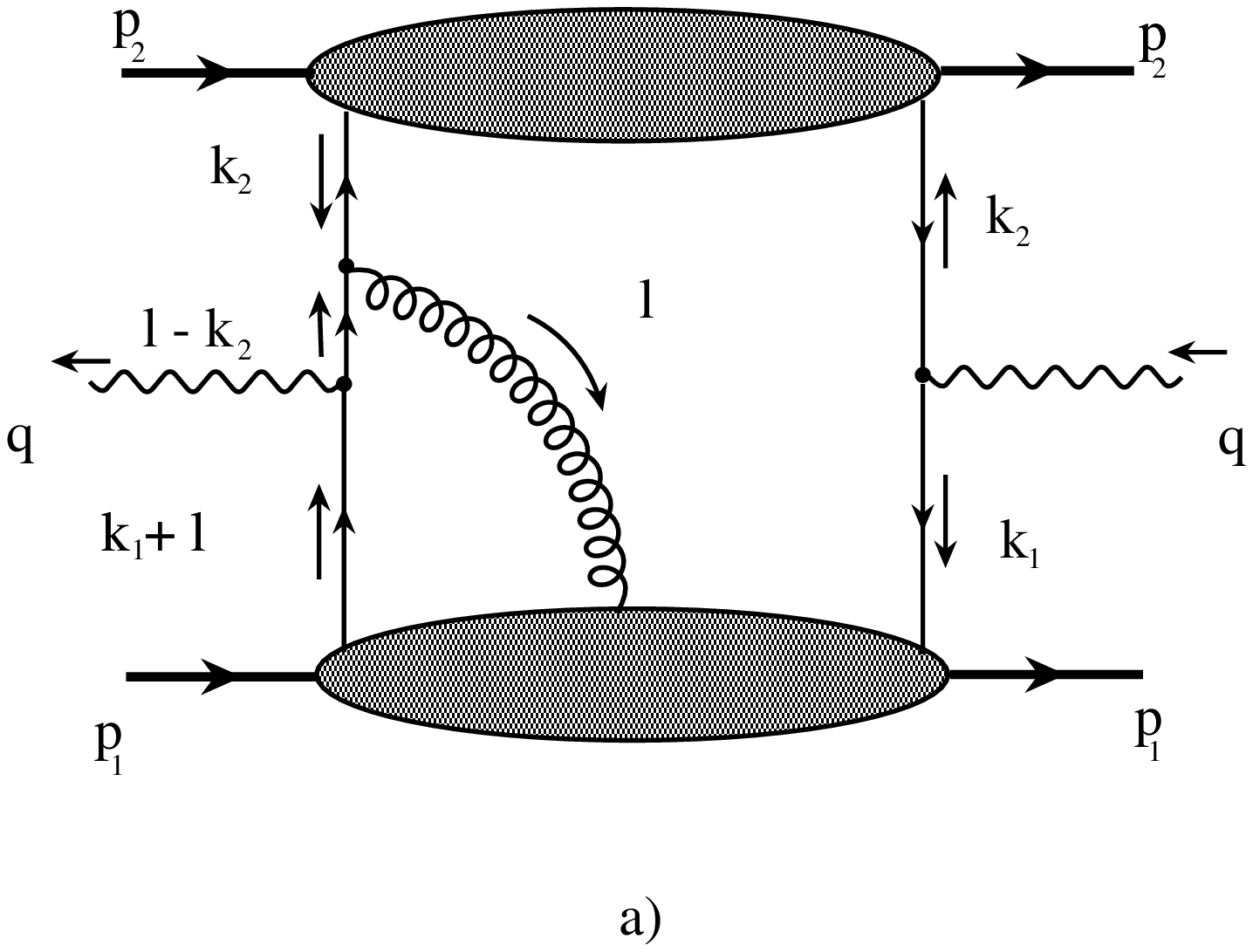}
\hspace{1.cm}\includegraphics[width=0.3\textwidth]{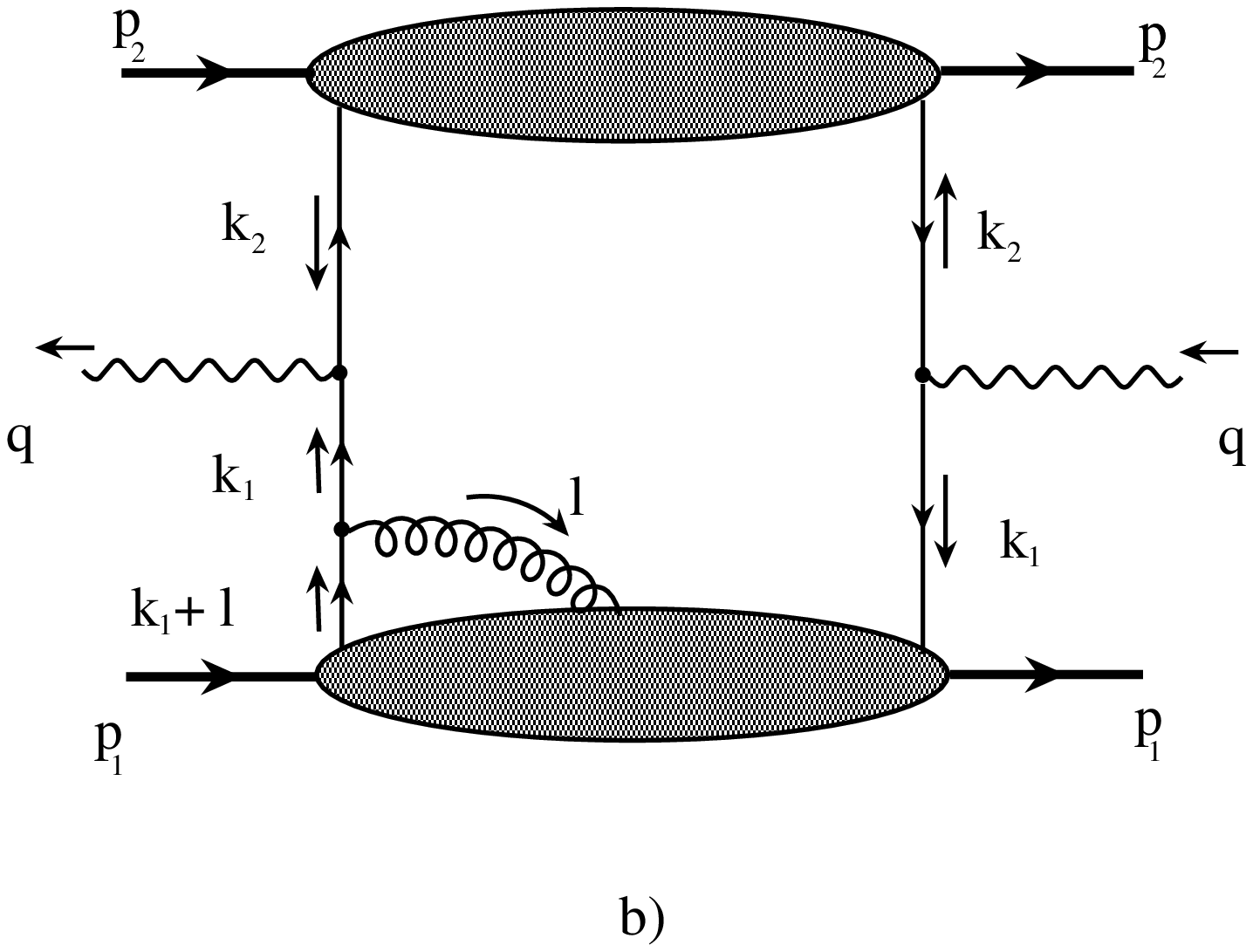}}
  \caption{The Feynman diagrams which contribute to the polarized Drell-Yan hadron tensor.}
\label{Fig-DY}
\end{figure}

\textbf{Causality and contour gauge for the gluonic pole}.
We study the contribution to the hadron tensor which is related to the single spin
(left-right) asymmetry
measured in the Drell-Yan process with the transversely polarized nucleon.
The DY process with the transversely polarized target
manifests \cite{Teryaev} the gluonic pole contributions.
Since we perform our calculations within a {\it collinear} factorization,
it is convenient (see,e.g., \cite{An})
to fix  the dominant light-cone directions for the DY process
shown at Fig. \ref{Fig-DY}
$p_1\approx Qn^{*+}/(x_B \sqrt{2}), \, p_2\approx Qn^-/(y_B \sqrt{2})$.
Focusing on the Dirac vector projection, containing the gluonic pole,
let us start with the standard hadron tensor
generated by the diagram depicted on Fig. \ref{Fig-DY}(a):
\begin{eqnarray}
\label{HadTen1-2}
{\cal W}^{(1)}_{\mu\nu}=\int d^4 k_1\, d^4 k_2 \, \delta^{(4)}(k_1+k_2-q)
\int d^4 \ell \,
\Phi^{(A)\,[\gamma^+]}_\alpha \, \bar\Phi^{[\gamma^-]}
\text{tr}\biggl[
\gamma_\mu  \gamma^- \gamma_\nu \gamma^+ \gamma_\alpha
\frac{\ell^+\gamma^- - k_2^-\gamma^+ -
\ell_T\gamma_T}
{-2\ell^+ k_2^- - \ell^2_T + i\epsilon}
\biggr] ,
\end{eqnarray}
where $\Phi^{(A)\,[\gamma^+]}_\alpha$ and $\bar\Phi^{[\gamma^-]}$
defined as in \cite{AT-GI-DY}.
Analyzing the $\gamma$-structure, {\it i.e}
$\gamma^+\gamma^\alpha\gamma^{\pm}$ in the trace, we may conclude that
(i) the $\ell^+\gamma^-$ term singles out
$\gamma^+ \gamma^\alpha \gamma^-$ with $\alpha=T$ which
will lead to $\langle \bar\psi\, \gamma^+ A^T_\alpha \psi\rangle$ giving
the contribution to SSA;
(ii) the $k_2^-\gamma^+$ term separates out
$\gamma^+ \gamma^\alpha \gamma^+$ with $\alpha=-$.
Therefore, this term will give $\langle \bar\psi\, \gamma^+\,  A^+\,\psi\rangle$
which will be exponentiated in the Wilson line
$[-\infty^-,\, 0^-]$;
(iii) the $\ell_T\gamma_T$ term separates out
$\gamma^+ \gamma^\alpha \gamma_T$ with $\alpha=T$ and, then,
will be exponentiated in the Wilson line
$[-\infty^-,-\infty_T\,;-\infty^-, 0_T]$.
Indeed, integrating over $\ell^+$, the $k_2^-$-term contribution
reads
\begin{eqnarray}
\label{HadTen1-4}
{\cal W}^{(1)\,[k_2^-]}_{\mu\nu}=\int d\mu (k_i;x_1,y)
\text{tr}\biggl[
\gamma_\mu  \gamma^- \gamma_\nu \gamma^+
\biggr] \bar\Phi^{[\gamma^-]}(k_2)
\int d^4\eta_1 e^{-ik_1\cdot\eta_1}
\langle \bar\psi(\eta_1)\gamma^+
ig \int\limits_{-\infty}^{+\infty} dz^- \theta(-z^-) A^+(z^-)
\psi(0)\rangle \, .
\end{eqnarray}
Including all gluon emissions from the antiquark going from the upper blob
on Fig. \ref{Fig-DY}(a), the $k_2^-$-type terms result in the
following matrix element:
\begin{eqnarray}
\label{me-Pexp}
\int d^4\eta_1 \, e^{-ik_1\cdot\eta_1}
\langle p_1, S^T | \bar\psi(\eta_1)\, \gamma^+ [-\infty^-,\, 0^-]
\psi(0) |S^T, p_1\rangle, \quad
[-\infty^-,\, 0^-]={\rm Pexp}\biggl\{ i g \int\limits^{-\infty}_{0}
 dz^- \, A^+(0,z^-,\vec{{\bf 0}}_T) \biggr\}.
\end{eqnarray}
If we include the mirror contributions, we will obtain
\begin{eqnarray}
\langle p_1,S_T| \bar\psi(\eta_1)\gamma^+ [\eta_1^-,-\infty^-] \psi(0)
|S_T,p_1 \rangle
\end{eqnarray}
which will ultimately give us the Wilson line connecting
the points $0$ and $\eta_1$.
This is exactly what happens in the spin-averaged DY process.
However, for the SSA we are interested in, these
two (direct  and mirror) diagrams have to be considered individually.
Their contributions to SSAs differ in sign and the dependence on
the boundary point at $-\infty^-$ does {\it not} cancel.
To eliminate the unphysical gluons from our consideration and use the
factorization scheme \cite{Efr},
we may choose a {\it contour} gauge \cite{ContourG}
\begin{eqnarray}
\label{cg2}
[-\infty^-,\, 0^-]=1 \,
\end{eqnarray}
which actually implies also the axial gauge $A^+=0$ used in \cite{Efr}.
Imposing this gauge one arrives \cite{ContourG} at the following representation
of the gluon field in terms of the strength tensor:
\begin{eqnarray}
\label{Ag}
A^\mu(z)=
\int\limits_{-\infty}^{\infty} d\omega^- \theta(z^- - \omega^-) G^{+\mu} (\omega^-)
+ A^\mu(-\infty) \, .
\end{eqnarray}
Moreover, if we choose instead an
alternative representation for the gluon in the form with $A^\mu(\infty)$,
keeping the causal prescription
$+i\epsilon$ in (\ref{HadTen1-2}), the cost of this will be the breaking of the electromagnetic gauge
invariance for the DY tensor.
Consider now the term with $\ell^+\gamma^-$ in (\ref{HadTen1-2})
which gives us finally the matrix element of the twist $3$
operator with the transverse gluon field.
The parametrization of the relevant matrix elements is
\begin{eqnarray}
\label{parVecDY}
&&\langle p_1, S^T | \bar\psi(\lambda_1 \tilde n)\, \gamma_{\beta} \,
g A_{\alpha}^T(\lambda_2\tilde n) \,\psi(0)
|S^T, p_1 \rangle
\stackrel{{\cal F}_2^{-1}}{=}
i\varepsilon_{\beta\alpha S^T p_1} \, B^V(x_1,x_2)\, .
\end{eqnarray}
Using the representation (\ref{Ag}), this function can be expressed as
\begin{eqnarray}
\label{Sol-way-1}
B^V(x_1,x_2)= \frac{T(x_1,x_2)}{x_1-x_2+i\epsilon} + \delta(x_1-x_2) B^V_{A(-\infty)}(x_1)\, ,
\end{eqnarray}
where the real regular function $T(x_1,x_2)$ ( $T(x,x) \neq 0$)
parametrizes
the vector matrix element of the operator involving the tensor $G_{\mu\nu}$ (cf. \cite{An-ImF}):
\begin{eqnarray}
\label{parT}
\langle p_1, S^T | \bar\psi(\lambda_1 \tilde n)\, \gamma_{\beta} \,
g\tilde n_\nu G_{\nu\alpha}(\lambda_2\tilde n) \,\psi(0)
|S^T, p_1 \rangle\stackrel{{\cal F}^{-1}_2}{=}
\varepsilon_{\beta\alpha S^T p_1}\,
T(x_1,x_2)\, .
\end{eqnarray}
Owing to the time-reversal invariance, the function $B^V_{A(-\infty)}(x_1)$,
\begin{eqnarray}
\label{BatInfty}
i\varepsilon_{\beta\alpha S^T p_1} \,\delta(x_1-x_2) B^V_{A(\pm\infty)}(x_1) \stackrel{{\cal F}}{=}
\langle p_1, S^T | \bar\psi(\lambda_1 \tilde n)\, \gamma_\beta \,
g A_{\alpha}^T(\pm\infty) \,\psi(0) | S^T, p_1 \rangle\, ,
\end{eqnarray}
can be chosen as
$B^V_{A(-\infty)}(x)= 0$.
Indeed, the function $B^V(x_1,x_2)$ is an antisymmetric function of its arguments \cite{Efr},
while the anti-symmetrization of the additional term with $B^V_{A(-\infty)}(x_1)$ gives zero.
If the only source of the imaginary part of the hadron tensor
is the quark propagator, one may realize this property by assumption:
$B^V(x_1,x_2)= T(x_1,x_2){\cal P}/(x_1-x_2)$
corresponding to
asymmetric boundary condition for gluons \cite{Boer}:
$B^V_{A(\infty)}(x) = - B^V_{A(-\infty)}(x)$.
Here we suggest another way of reasoning. The causal prescription for the quark propagator,
generating its imaginary part, simultaneously leads to the imaginary part of the gluonic pole.
We emphasize that this does not mean the appearance of imaginary part of matrix element
but rather the prescription of its convolution with hard part (see e.g. \cite{Rad}).
Note that the fixed complex prescription $+i\epsilon$ in the
gluonic pole of $B^V(x_1,x_2)$ (see, (\ref{Sol-way-1})) is one of our main results and
is very crucial for an extra contribution to hadron tensor we are now ready to explore.
Indeed, the gauge condition must be the same for all the diagrams, and
it leads to the appearance of imaginary phase of the diagram (see, Fig. \ref{Fig-DY}(b))
which naively does not have it. Let us confirm this by explicit calculation.

\textbf{Hadron tensor and gauge invariance}.
We now return to the hadron tensor and calculate the part involving $\ell^+\gamma^-$,
obtaining the following expression for
the standard hadron tensor (see, the diagram on Fig. \ref{Fig-DY}(a)):
\begin{eqnarray}
\label{FacHadTen1}
\overline{\cal W}^{(1)\,[\ell^+]}_{\mu\nu}=
- \bar q(y_B)
\Im m\, \int dx_2 \, \text{tr}\biggl[
\gamma_\mu \gamma_\beta \gamma_\nu \hat p_2 \gamma^T_\alpha
\frac{(x_B-x_2)\hat p_1}{(x_B-x_2)ys + i\epsilon}
\biggr] B^V(x_B,x_2) \, \varepsilon_{\beta\alpha S^T p_1} \, .
\end{eqnarray}
We are now in position to check the QED gauge invariance by contraction
with the photon momentum $q_\mu$.
Calculating the trace, one gets
\begin{eqnarray}
\label{FacHadTen4}
q_\mu \, \overline{\cal W}^{(1)}_{\mu\nu}= - \bar q(y_B)
\varepsilon_{\nu p_2 S^T p_1}\, \int\limits_{-1}^{1} dx_2\, \Im m
\frac{x_B-x_2}{x_B-x_2+i\epsilon} B^V(x_B,x_2) \not= 0\, ,
\end{eqnarray}
if the gluonic pole is present.
We now focus on the contribution from the diagram depicted on Fig. \ref{Fig-DY}(b).
The corresponding hadron tensor takes the form:
\begin{eqnarray}
\label{HadTen2}
{\cal W}^{(2)}_{\mu\nu}=
\int d^4 k_1\, d^4 k_2 \, \delta^{(4)}(k_1+k_2-q)
\text{tr}\biggl[
\gamma_\mu  {\cal F}(k_1) \gamma_\nu \bar\Phi(k_2)
\biggr]
\, ,
\end{eqnarray}
where the function ${\cal F}(k_1)$ reads
\begin{eqnarray}
\label{PhiF2}
{\cal F}(k_1)= S(k_1) \gamma_\alpha \int d^4\eta_1\, e^{-ik_1\cdot\eta_1}
\langle p_1, S^T | \bar\psi(\eta_1) \, gA^T_{\alpha}(0) \, \psi(0) |S^T, p_1\rangle \, .
\end{eqnarray}
Performing the collinear factorization, we derive the expression for the
factorized hadron tensor which corresponds to the diagram on Fig. \ref{Fig-DY}(b):
\begin{eqnarray}
\label{FacHadTen2}
\overline{\cal W}^{(2)}_{\mu\nu}=  \int dx_1 \, dy \,
\biggl[\delta(x_1-x_B) \delta(y-y_B)\biggr] \, \bar q(y) \,
\text{tr}\biggl[
\gamma_\mu \biggl( \int d^4 k_1\,
\delta(x_1p_1^+ - k_1^+) {\cal F}(k_1)\biggr) \gamma_\nu \hat p_2 \biggr] \, .
\end{eqnarray}
After some algebra, the integral over $k_1$ in (\ref{FacHadTen2}) can be
rewritten as
\begin{eqnarray}
\label{FacF2}
\int d^4 k_1\, \delta(x_1p_1^+ - k_1^+) {\cal F}^{[\gamma^+]}(k_1)=
\frac{\hat p_2 \gamma_\alpha^T\gamma_\beta}{2p_2^-p_1^+}
\, \varepsilon_{\beta\alpha S^T p_1}\,
\frac{1}{x_1+i\epsilon} \,
\int\limits_{-1}^{1} dx_2\, B^V(x_1,x_2)\, ,
\end{eqnarray}
where the parametrization (\ref{parVecDY}) has been used.
Taking into account (\ref{FacF2}) and calculating the Dirac trace, the
contraction of the tensor $\overline{\cal W}^{(2)}_{\mu\nu}$ with the
photon momentum $q_\mu$ gives us
\begin{eqnarray}
\label{FacHadTen3}
q_\mu \, \overline{\cal W}^{(2)}_{\mu\nu}= \int dx_1 \, dy \,
\biggl[\delta(x_1-x_B) \delta(y-y_B)\biggr] \,
\bar q(y) \,
 \, \varepsilon_{\nu p_2 S^T p_1}\,
 \int\limits_{-1}^{1} dx_2\, \Im m \,B^V(x_1,x_2)\, .
\end{eqnarray}
If the function $B^V(x_1,x_2)$ is the purely
real one,
this part of the hadron tensor does not contribute to the imaginary part.
We now study the $\overline{\cal W}^{(1)}_{\mu\nu}$ and
$\overline{\cal W}^{(2)}_{\mu\nu}$ contributions and its role for the QED gauge invariance.
One can easily obtain:
\begin{eqnarray}
\label{com}
q_\mu \, \overline{\cal W}^{(1)}_{\mu\nu} + q_\mu \, \overline{\cal W}^{(2)}_{\mu\nu}=
\varepsilon_{\nu p_2 S^T p_1}\, \bar q(y_B)\,
 \Im m\, \int\limits_{-1}^{1} dx_2\, B^V(x_B,x_2)\,
 \biggl[  \frac{x_B-x_2}{x_B-x_2+i\epsilon} -  1 \biggr]\, .
\end{eqnarray}
Assuming the gluonic pole in $B^V(x_1,x_2)$ exists, after
inserting (\ref{Sol-way-1}) into (\ref{com}), one gets
\begin{eqnarray}
\label{com-3}
q_\mu\, \overline{\cal W}^{(1)}_{\mu\nu} +
q_\mu\, \overline{\cal W}^{(2)}_{\mu\nu} = 0\, .
\end{eqnarray}
This is nothing else than the QED gauge invariance for the imaginary part of the hadron tensor.
We can see that the gauge invariance takes place only if the prescriptions
in the gluonic pole and in the quark propagator of the hard part are coinciding.
As we have shown, only the sum of two contributions represented by the diagrams on
Fig. \ref{Fig-DY}(a) and (b) can ensure the electromagnetic gauge invariance.
We now inspect the influence of a ``new" contribution \ref{Fig-DY}(b) on the single spin asymmetry
and obtain the QED gauge invariant expression for the hadron tensor.
It reads
\begin{eqnarray}
\label{HadTen-GI}
\overline{\cal W}^{GI}_{\mu\nu}=
\overline{\cal W}^{(1)}_{\mu\nu} + \overline{\cal W}^{(2)}_{\mu\nu} =
- \frac{2}{q^2}\,\varepsilon_{\nu S^T p_1 p_2} \,
[x_B \, p_{1\,\mu}- y_B\, p_{2\,\mu}]
\, \bar q(y_B)\, T(x_B,x_B) \, .
\end{eqnarray}
Within the lepton c.m. system, the SSA \cite{Teryaev} related to the gauge invariant
hadron tensor (\ref{HadTen-GI}) reads
\begin{eqnarray}
{\cal A}^{SSA} =
2\, \frac{\cos\phi \, \sin 2\theta\, T(x_B,x_B)}{M (1+\cos^2\theta) q(x_B)} ,
\end{eqnarray}
where $M$ is the dilepton mass.
We want to emphasize that this differs by the factor of $2$ in comparison with the case where
only one diagram, presented on Fig. \ref{Fig-DY}(a), has been included in the (gauge non-invariant)
hadron tensor.
Therefore, from the practical point of view, the neglecting of the diagram on
Fig. \ref{Fig-DY}(b) or, in other words,
the use of the QED gauge non-invariant hadron tensor yields the error of the factor of two.

\textbf{Conclusions and Discussions}.
Shortly summarizing, we want to notice that
if we start to work within the axial (light-cone) gauge, without any referring to
the contour gauge,
we have to sort out all possible prescriptions in order to choose such ones which are in agreement
with the gauge invariance \cite{BIS}.
Also, if we ``blindly" work within the axial (light-cone) gauge, in order to
get the gauge invariance, we are forced to introduce a such specific subject as the
so-called special propagator {\it a la} J.w.Qiu. \cite{BQ,Cao:2009rq}.
On the other hand, having considered the axial gauge as a particular case of the path-dependent
contour gauge, we have no ambiguities with the prescriptions which automatically agree with
the gauge invariance. The practical issue of that the gauge invariance has been restored is
the new-found factor of $2$ in the expression for SSA in
the transverse polarized Drell-Yan process.

Thus, we showed that it is mandatory to include
a contribution of the extra diagram which naively does not have an imaginary part.
The account for this extra contribution
leads to the amplification of SSA by the factor of $2$.
This additional contribution emanates
from the complex gluonic pole prescription in the representation of the twist $3$ correlator
$B^V(x_1,x_2)$ which, in its turn,
is directly related to the complex pole prescription in the quark propagator forming
the hard part of the corresponding hadron tensor.
The causal prescription in the quark propagator, involved in the hard part of
the diagram on Fig.\ref{Fig-DY}(a), selects from the physical axial gauges the contour gauge.
We argued that, in addition to the electromagnetic gauge invariance,
the inclusion of new-found contributions corrects by the factor of $2$ the expression for SSA in
the transverse polarized Drell-Yan process.
We proved that the complex prescription in the quark propagator
forming the hard part of the hadron tensor,
the starting point in the contour gauge,
the fixed representation of $B^V(x_1,x_2)$
and the electromagnetic gauge invariance of the hadron tensor must be considered together
as the deeply related items.
In recent work \cite{MZh}, the factor $1/2$ instead of $2$ has been claimed.
So, in addition to the sign puzzle, do we have a factor of $2$ puzzle?!

\section{Acknowledgements}

This work is supported in part by
the RFBR (grants 12-02-00613).


\end{document}